\documentclass[journal,twoside,web]{ieeecolor}
\usepackage{generic}
\usepackage{cite}
\usepackage{amsmath,amssymb,amsfonts}
\usepackage{algorithmic}
\usepackage{graphicx}
\usepackage{textcomp}
\usepackage[hidelinks]{hyperref} 
\usepackage{cuted} 
\def\BibTeX{{\rm B\kern-.05em{\sc i\kern-.025em b}\kern-.08em
    T\kern-.1667em\lower.7ex\hbox{E}\kern-.125emX}}
\begin{document}

\title{Stratified Multivariate Multiscale Dispersion Entropy for Physiological Signal Analysis}
\author{Evangelos Kafantaris, Tsz-Yan Milly Lo, and Javier Escudero 
\thanks{The work of Evangelos Kafantaris was supported in part by the Engineering and Physical Sciences Research Council, the National Health Service of the United Kingdom, and the Network of European Funding for Neuroscience Research, which is part of the European Research Area Networks, through the funding of a Ph.D. Studentship. The work of Javier Escudero was supported by the Leverhulme Trust via a Research Project under Grant RPG-2020-158. (Corresponding author: Evangelos Kafantaris.) Evangelos Kafantaris is with the School of Engineering, Institute for Digital Communications, University of Edinburgh, Edinburgh EH9 3FB, U.K. (e-mail: evangelos.kafantaris@ed.ac.uk). Tsz-Yan Milly Lo is with the Centre of Medical Informatics, Usher Institute, University of Edinburgh, U.K., and also with the Royal Hospital for Children and Young People, U.K. Javier Escudero is with the School of Engineering, Institute for Digital Communications, University of Edinburgh, U.K. 
}
}

\maketitle

\begin{abstract}
Multivariate entropy quantification algorithms are becoming a prominent tool for the extraction of information from multi-channel physiological time-series. However, in the analysis of physiological signals from heterogeneous organ systems, certain channels may overshadow the patterns of others, resulting in information loss. Here, we introduce the framework of Stratified Entropy to prioritize each channels' dynamics based on their allocation to respective strata, leading to a richer description of the multi-channel time-series. As an implementation of the framework, three algorithmic variations of the Stratified Multivariate Multiscale Dispersion Entropy are introduced. These variations and the original algorithm are applied to synthetic time-series, waveform physiological time-series, and derivative physiological data. Based on the synthetic time-series experiments, the variations successfully prioritize channels following their strata allocation while maintaining the low computation time of the original algorithm. In experiments on waveform physiological time-series and derivative physiological data, increased discrimination capacity was noted for multiple strata allocations in the variations when benchmarked to the original algorithm. This suggests improved physiological state monitoring by the variations. Furthermore, our variations can be modified to utilize a priori knowledge for the stratification of channels. Thus, our research provides a novel approach for the extraction of previously inaccessible information from multi-channel time series acquired from heterogeneous systems.

\end{abstract}


\section{Introduction}
\label{sec:introduction}

\IEEEPARstart{I}{ncreased} amounts of physiological data are becoming available due to the advances in physiological recording technology across a range of applications from wearable devices to clinical environments \cite{yetisen_wearables_2018,witt_windows_2019,paine_systematic_2016}. The analysis of these data can contribute to effective prognosis, early stage intervention, personalised treatments, and improved clinical decision making. However, for the successful development of algorithms capable of extracting viable information, certain characteristics of the data have to be considered. These include their multivariate nature due to the interaction of multiple organ systems in human physiology \cite{cerutti_multivariate_2012,pereda_nonlinear_2005,cerutti_multiscale_2009,bartsch_network_2015,fossion_physicists_2018}, the potential non-linear nature of their dynamics \cite{goldberger_fractal_2002,goldberger_what_2002,scafetta_fractal_2007,martis_application_2011,pradhan_higher-order_2012,muller_causality_2016}, and the low data-quality arising from the recording conditions \cite{azimi_missing_2019,kumar_automated_2016,moody_physionet/computing_2010}.

Entropy quantification algorithms are becoming a prominent tool for the  measurement of dynamics from uni- and multi-channel time-series \cite{ribeiro_entropy_2021}. These algorithms can be broadly characterised into those based on Shannon Entropy \cite{shannon_mathematical_1948} or on Conditional Entropy defined as the quantity of information observed in a sample at a time-point $n$ that cannot be explained based on previous samples up to time point $n-1$ \cite{papoulis_probability_2002}. They have been successful in a variety of applications such as the monitoring of machine operation \cite{rostaghi_application_2019,huo_entropy_2020} and the analysis of financial time-series \cite{wu_modified_2018,zhang_multivariate_2019}.

The quantification of entropy -- as a measure of physiological signals' complexity -- is of direct interest for the monitoring of a system's physiological states, particularly when considering the Critical Slow Down (CSD) and Loss of Complexity (LoC) paradigms as well as their combination within the scope of the ``entropy pump'' (EP) hypothesis. The CSD paradigm considers that during frail or pathological states, a slowing down is observed in the capacity of the system to recover from external stressors resulting in increased output complexity for certain regulatory variables  \cite{scheffer_early-warning_2009, olde_rikkert_slowing_2016,nakazato_estimation_2020}. The LoC paradigm suggests that when the equilibrium of a system is disrupted, multiple processes that displayed multi-scale complexity produce output measurements of reduced complexity indicating a loss in the system's flexibility and capacity to adapt in the presence of external stressors \cite{lipsitz_dynamics_2002,goldberger_what_2002}. These seemingly opposite paradigms are combined in the EP hypothesis that separates physiological parameters in regulated and effector variables. Based on this hypothesis, an ``entropy pump'' is observed, thanks to which homeostasis is achieved, by maintaining a stable, low complexity output for regulated variables through the complex and variable outputs of effector variables \cite{fossion_physicists_2018,barajas-martinez_physiological_2022}. A pathological state is observed when its direction is disrupted and an increase in the complexity of regulated variables is observed as per the CSD paradigm, while a decrease in the complexity of the effector variables is observed in accordance with the LoC paradigm.

Entropy has been extensively used to analyse physiological signals. Examples include algorithms based on Shannon Entropy such as Permutation Entropy (PEn) \cite{bandt_permutation_2002} and Dispersion Entropy (DisEn) \cite{rostaghi_dispersion_2016,azami_amplitude-_2018} to analyse electoencephalogram (EEG) signals to track the state of consciousness of patients under the effect of anaesthetic drugs \cite{olofsen_permutation_2008}, and to analyse blood pressure signals to quantify the effect of aging in the reduction of the recorded signal's variability \cite{rostaghi_dispersion_2016}, respectively. Algorithms based on Conditional Entropy have also been utilized, such as Approximate Entropy (ApEn) \cite{pincus_approximate_1991} for the investigation of abnormalities in respiratory function caused by panic disorders \cite{caldirola_approximate_2004}, Sample Entropy (SampEn) \cite{richman_physiological_2000} for the analysis of neonatal heart rate variability to diagnose sepsis \cite{lake_sample_2002}, and Fuzzy Entropy (FuzzyEn) on surface electromyography (EMG) signals for the detection of motion \cite{weiting_chen_characterization_2007}.

For the effective analysis of physiological dynamics, multi-channel time-series have to be analyzed both in a univariate and a multivariate manner. This is a necessary step to ensure that, cross-channel dynamics can be quantified to allow the study of dynamics developed across different components of the same organ system as well as across distinct  systems \cite{cerutti_multiscale_2009,cerutti_multivariate_2012,pereda_nonlinear_2005,bartsch_network_2015,fossion_physicists_2018}. For this reason, recent research has focused on producing multivariate variations of entropy algorithms to extract features from two or more channels: DisEn \cite{azami_multivariate_2019}, PEn \cite{morabito_multivariate_2012}, ApEn, SampEn \cite{ahmed_multivariate_2011}, and FuzzyEn \cite{azami_memd-enhanced_2016}. 

However, while multivariate algorithms can extract an output feature from a multi-channel time-series, the approach is limited with regards to the total information retrieved. The dynamics of certain input channels may overshadow those of others due to the potentially different dominant frequencies amongst the physiological signals of each channel. This becomes apparent when multi-channel time-series are comprised of signals that arise from heterogeneous organ systems such as the combination of electrocardiograms (ECG) \cite{luo_review_2010}, EEG \cite{dressler_awareness_2004}, arterial blood pressure (BP) \cite{stauss_identification_2007}, and nasal respiratory (RESP) signals \cite{kaczka_assessment_1995,diong_modeling_2007}, whose dominant frequencies and temporal structures display clear differences.

As a step towards addressing this challenge, recent studies have suggested non-uniform multiscale embedding between the input channels. This approach aims to find the optimal combination of scales for the analysis of the multi-channel time-series so that each channel is analyzed at the scale where most of its dynamics would arise, by modifying the time-delay \cite{gu_optimizing_2022} or the embedding dimension used for each channel \cite{xiao_variational_2021}. While this approach offers an interesting and modular configuration of analysis, it faces challenges that limit its applicability. These are the potential mismatch of each channel's data length with the optimal scale values, the limitation of multiscale analysis to specific scales for each channel resulting in an incomplete multiscale output, the instability of the method for increased number of channels, and the potential for overshadowing to occur even at optimal scale combinations.

A different approach for the analysis of interdependencies within a group of multi-channel time-series arises from the utilization of Cross-Entropy algorithms, developed for ApEn, SampEn \cite{richman_physiological_2000}, FuzzyEn \cite{xie_cross-fuzzy_2010}, and PEn \cite{shi_coupling_2015}. With them, an entropy based feature quantifies the coupling between two channels. The variations of SampEn and FuzzyEn are non-directional, while the variations of ApEn and PEn are directional. In the latter cases, one of the two channels acts as a ``designated'' channel in the measurement. Thus, the potential overshadowing of each channels' dynamics could be avoided since each channel has the opportunity to be designated. However, this approach is limited to bivariate measurements between two channels. Therefore, it cannot capture higher-order dynamics arising jointly from three or more channels. A second limitation is that, by definition, it measures the coupling between the two channels and is not a measurement of their combined dynamics.

In this study, we propose the framework of Stratified-Entropy to combine positive elements of both the Multivariate and Cross- Entropy algorithms to increase the information that can be extracted from a set of multi-channel time-series by allowing each channel's dynamics to have a different level of prioritization during the quantification of the output entropy value based on its allocation to a respective stratum. Namely, the main contributions of the presented work are:

\begin{itemize}

\item The introduction of  the  Stratified-Entropy framework as a new form of multivariate and multiscale analysis that increases the amount of information extracted from a multi-channel time-series via entropy quantification algorithms.  

\item  The implementation of the Stratified-Entropy framework through the introduction of three novel algorithms of Stratifed Multivariate Multiscale Dispersion Entropy (SmvMDE) that prioritize channels during the calculation of the output entropy value based on their allocation to hierarchical strata.

\item The analysis and benchmarking of the SmvMDE algorithms through experiments applied to synthetic time-series, waveform physiological time-series, and derivative physiological data. 

\end{itemize}

\section{Methods}

\subsection{Stratified Entropy Framework}

Within the framework of Stratified Entropy, strata are defined with a clear hierarchy of prioritization. The number of strata can vary based on the implementation of the framework. Each channel is allocated to one of the available strata and every channel has a weighted contribution in the calculation of the output entropy feature based on their allocated stratum. 

We build upon the existing Multivariate Multiscale Dispersion Entropy (mvMDE) algorithm \cite{azami_multivariate_2019} and introduce three novel variations of the Stratified Multiscale Multivariate Dispersion Entropy algorithm (SmvMDE): The Threshold (T-SmvMDE), Soft Threshold (ST-SmvMDE), and Proportional (P-SmvMDE) variations.

For the purposes of this study, all three variations have been designed based on a two strata configuration, a \textit{core} stratum (prioritized) and a \textit{periphery} stratum. The potential extension to configurations with higher numbers of strata is discussed in Subsection~\ref{Strata_Number}. The variations differ in how the dynamics of channels allocated to the core stratum are prioritised over the periphery channels.

The following subsections start with a description of the original mvMDE algorithm, continue with the introduction of the SmvMDE variations and the changes they introduce to mvMDE, and describe the experiments conducted to analyse and benchmark their operation.

\subsection{Multivariate Multiscale Dispersion Entropy}
\label{MDE}
DisEn arises from the integration of Shannon Entropy with symbolic dynamics, aiming to quantify the degree of irregularity in an input time-series segment. It achieves good discrimination capacity between different types of physiological activity while maintaining a low computational time \cite{rostaghi_dispersion_2016,azami_amplitude-_2018}.
The mvMDE algorithm allows the multivariate quantification of DisEn from multi-channel time-series taking into consideration both temporal and spatial dynamics, across multiple time scales \cite{azami_multivariate_2019}. 

\subsubsection{Coarse-Graining Process for Multiscale Implementation}

For the successful quantification of a time-series' complexity across multiple time scales, a number of coarse graining procedures have been suggested. These include the widely used moving average approach \cite{costa_multiscale_2002,ahmed_multivariate_2012,valencia_refined_2009}, low-pass Butterworth filtering \cite{valencia_refined_2009,hamed_azami_coarse-graining_2018}, and empirical mode decomposition \cite{hamed_azami_coarse-graining_2018}. This study builds upon the original algorithmic implementation of mvMDE and therefore utilizes the moving average coarse graining approach for simplicity \cite{azami_multivariate_2019}, although other alternatives provide better frequency responses. Based on this approach, in a set of $p$-channel time-series $\mathbf {Y} =\{y_{k,b}\}_{k=1,2,\cdots,p}^{b=1,2,\cdots,N}$, each channel is processed separately and divided into non-overlapping segments of length equal to the defined time scale factor, $\tau$. For each segment, an average value is calculated and used to derive the coarse-grained multi-channel time-series as follows:
\begin{equation}
x_{k,i}(\tau) = \frac{1}{\tau} \sum_{b=(i-1)\tau +1}^{i\tau} {y_{k,b}} , 1 \leq i \leq \left \lfloor{{\frac{L}{\tau}}}\right \rfloor= N, 1 \leq k \leq p
\end{equation}
where $L$ is the original channel length and $N$ the resulting coarse-grained channel length.

\subsubsection{Application of Mapping Function}
For the implementation of mvMDE, a recommended step is the application of a non-linear mapping function to each channel, such as the normal cumulative distribution function (NCDF) \cite{azami_multivariate_2019}. The selection of a non-linear over a linear mapping function seeks to ensure that maximum and minimum amplitude values, that can be significantly larger or smaller than the mean value of the channel, do not disrupt the allocation of samples to classes by forcing the majority of samples to be assigned to a small number of classes \cite{rostaghi_dispersion_2016,azami_amplitude-_2018,kafantaris_augmentation_2020}. For multiscale implementations using NCDF, the mean and standard deviation of the original non coarse-grained time-series are used and remain constant for the mapping process across all temporal scale factors. This ensures that the mapping based on the NCDF remains fixed and is not affected by the averaging taking place during the coarse graining process \cite{azami_multivariate_2019}.

\subsubsection{Algorithm for mvMDE}
\label{mvMDE_steps}
For a set of $p$-channel time-series $\mathbf {X} =\{x_{k,i}\}_{k=1,2,\cdots,p}^{i=1,2,\cdots,N}$ of length $N$ each, the computational steps of mvMDE are the following \cite{azami_multivariate_2019}:

\begin{enumerate}

\item {Production of univariate quantised time-series:} A number of classes ($1,2,\ldots,c$) are distributed along the amplitude range of each channel separately. Their samples are allocated to their nearest respective class based on their amplitude. As a result, a quantized channel $u_j (j= 1,2,\dotsc,N)$ is produced for each respective input channel, resulting in a set of $p$-quantized channels $\mathbf {U} =\{u_{k,i}\}_{k=1,2,\cdots,p}^{i=1,2,\cdots,N}$. 

\item {Formulation of multivariate embedded vectors:} From $\{u_{k,i}\}$, the quantized samples are embedded into univariate vectors of length $m$ (with a time delay $d$) for each channel. The univariate embedded vectors are then combined in sets of $p$-synchronised vectors, one from each channel. The vectors within each synchronised set are serially concatenated for the production of a respective multivariate embedded vector $Z(j)$, of length $m\cdot p$, for each $j =1,2,\dotsc,N - (m-1)d$.

\item {Mapping to multiple dispersion patterns:} In mvMDE, each embedded vector is mapped to multiple dispersion patterns to effectively evaluate patterns both temporally within the same channel as well as across channels. Each subset of $m$ elements in $Z(j)$ is accessed, following all possible $\binom{m\cdot p}{m}$  combinations. This formulates $\phi_q(j) (q=1,\dots \binom{m\cdot p}{m})$ embedded subvectors that are then mapped to their corresponding ${{\pi }_{{{v}_{0}}\dots{{v}_{m-1}}}}$ dispersion pattern. As a result, the total number of dispersion pattern instances is $ (N - (m-1)d) \binom{m\cdot p}{m}$ and the number of unique dispersion patterns is $c^m$.

\item {Calculation of Dispersion Pattern Relative Frequency:} 
For each of the $c^m$ unique dispersion patterns, their relative frequency is calculated as follows, with $\#$ being the symbol that denotes the cardinality of the set:
\begin{equation}
\begin{gathered}
p({{\pi }_{{{v}_{0}}\dots{{v}_{m-1}}}})=\\ \frac{\#\{j\left| j\le N-(m-1)d,\phi_q(j)\text{ has type }{{\pi }_{{{v}_{0}}\dots{{v}_{m-1}}}} \right.\}}{(N-(m-1)d)\binom{mp}{m}}
\end{gathered}
\end{equation}

\item {Calculation of Multivariate Dispersion Entropy:}  Utilizing the  relative frequencies of the dispersion patterns considering both temporal and spatial domains as above, the output entropy value for $X$ is calculated based on Shannon's entropy and is normalized in the range of $0$ to $1$ by dividing with $\ln{c^m}$ : 
\end{enumerate}

\begin{equation}
\begin{split}
mvMDE(\mathbf{X},m,c,d)=-\sum_{\pi=1}^{c^m} {\frac{p({{\pi}_{{{v}_{0}}\dots{{v}_{m-1}}}})\cdot\ln \left(p({{\pi}_{{{v}_{0}}\dots{{v}_{m-1}}}}) \right)} {\ln{c^m}}}      
\end{split}
\end{equation}

\subsection{Stratified-Dispersion Entropy Variations}

Building on the original mvMDE, we introduce three variations of SmvMDE as implementations of the Stratified Entropy Framework. With their two strata configuration, the SmvMDE variations separate the channels in two sets. The set of one or more designated channels, which are allocated to the ``core'' stratum, and the set of secondary channels which are allocated to the ``periphery'' stratum. 

The original mvMDE treats all embedded subvectors as equal. Instead, the SmvMDE variations prioritise subvectors that contain samples retrieved from designated channels. The Threshold (T-SmvMDE), Soft Threshold (ST-SmvMDE), and Proportional (P-SmvMDE) variations use distinct approaches for adjusting the contribution of each combination by modifying the third and fourth steps of the original mvMDE algorithm, as described in Subsection~\ref{mvMDE_steps}. The code implementing the SmvMDE variations in Matlab is publicly available at: \url{https://github.com/EvangelosKafantaris/SmvMDE.git}.

\subsubsection{Threshold Variation}

\begin{figure}[!t]
\centerline{\includegraphics[width=1\columnwidth]{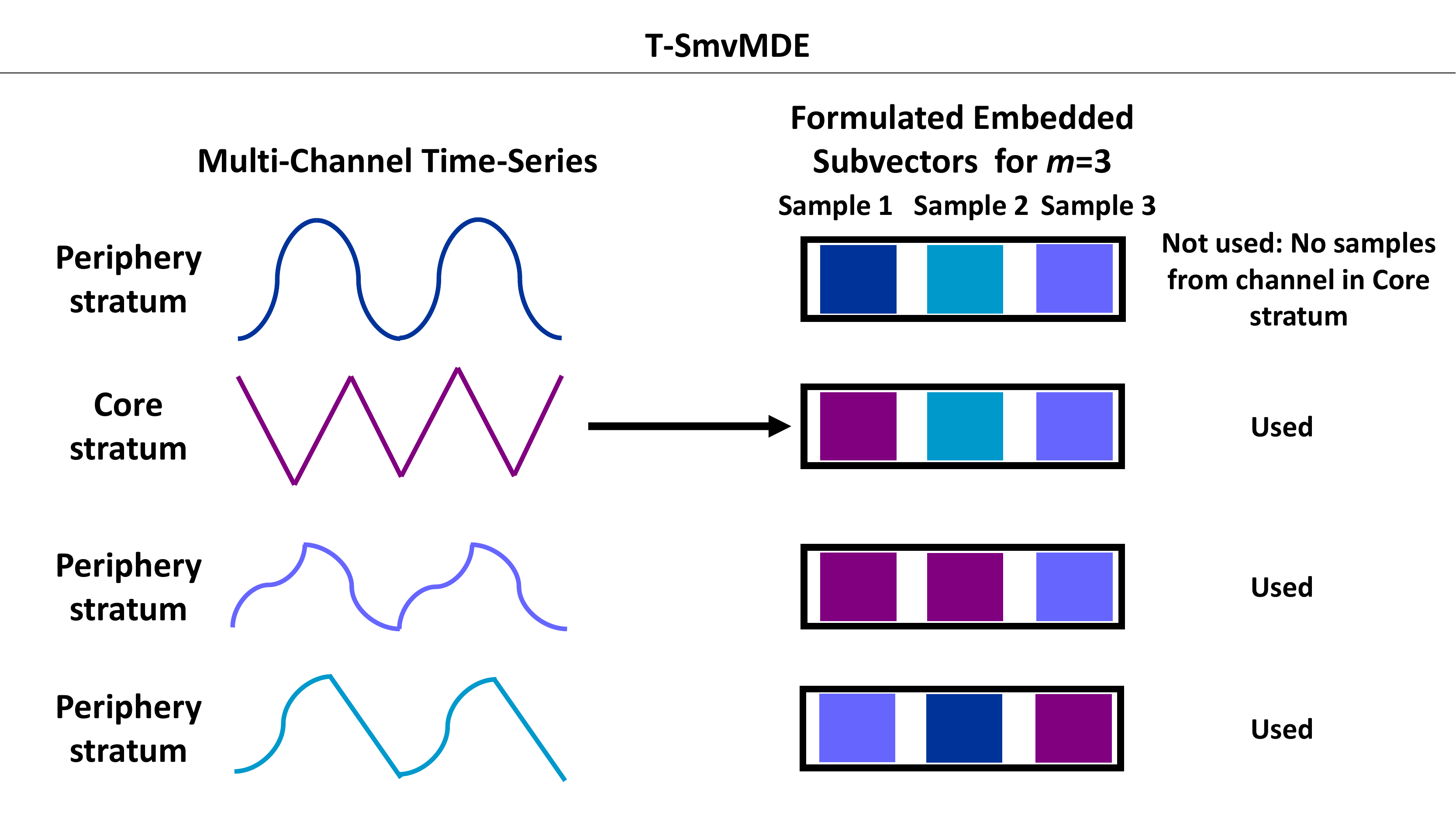}}
\caption{A representation of the embedded subvector inclusion and exclusion process for the T-SmvMDE variation with $m = 3$ and $t = 1$. }
\label{T_SmvMDE_Diagram}
\end{figure}

T-SmvMDE defines the minimum number of samples extracted from designated channels that each subvector should contain in order to be considered. This is achieved through a new input parameter: the threshold ($t$). The initially $\binom{m\cdot p}{m}$ subvectors utilized in the case of the original mvMDE are reduced to a subset of length $l_t$ that only includes subvectors that meet or surpass the threshold of having $t$ or more samples in the patterns of length $m$. As a result, for each multivariate embedded vector $Z(j)$ only $\phi_q(j) (q=1,\dots l_t)$ subvectors are mapped to dispersion patterns. This results in the reduction of dispersion pattern instances to $(N - (m-1)d) l_t$.  Fig.~\ref{T_SmvMDE_Diagram} displays a diagram illustrating the T-SmvMDE subvector selection process. 

For each unique dispersion pattern, their relative frequency is calculated with a modified denominator to match the reduced number of dispersion patterns:
\begin{equation}
\begin{gathered}
p({{\pi }_{{{v}_{0}}\dots{{v}_{m-1}}}})=\\ \frac{\#\{j\left| j\le N-(m-1)d,\phi_q(j)\text{ has type }{{\pi }_{{{v}_{0}}\dots{{v}_{m-1}}}} \right.\}}{(N-(m-1)d)l_t}
\end{gathered}
\end{equation}

\subsubsection{Soft Threshold Variation}
\label{ST_introduction}

As an intermediate algorithm between T-SmvMDE and mvMDE, ST-SmvMDE combines the $t$ input parameter with the additional reduced weight ($w$) parameter to reduce the contribution of subvectors that do not meet the threshold of $t$, without removing them completely. The possible values of the $w$ parameter range from a minimum value of $0$, where the output value will match that of T-SmvMDE, to a maximum value of $1$, where the output will match that of the original mvMDE algorithm, since no reduction of contribution will occur.

Based on $t$, the subvectors are split into two subsets: A primary subset with length $l_p$ whose contribution to the calculation of a dispersion pattern's frequency remains unchanged; and a secondary subset with length $l_s$ whose impact is reduced by multiplying the number of respective dispersion pattern instances with $w$. Consequently, for each $Z(j)$: $\phi_p(j) (p=1,\dots l_p)$ subvectors are formulated from the primary and $\phi_s(j) (s=1,\dots l_s)$ from the secondary subset, respectively.

Therefore, the maximum value of instances for a dispersion pattern becomes $(N - (m-1)d)(l_p+(l_s w))$. As a result, for each unique dispersion pattern, their relative frequency is:
\begin{equation}
\begin{gathered}
p({{\pi }_{{{v}_{0}}\dots{{v}_{m-1}}}}) =\\ \frac{\#\{j\left| j\le N-(m-1)d,\phi_p(j)\text{ has type }{{\pi }_{{{v}_{0}}\dots{{v}_{m-1}}}} \right.\}}{(N-(m-1)d)\ (l_p + l_s w)} \\ + \frac{\#\{j\left| j\le N-(m-1)d,\phi_s(j)\text{ has type }{{\pi }_{{{v}_{0}}\dots{{v}_{m-1}}}} \right.\}}{(N-(m-1)d)\ (l_p + l_s w)} \cdot w
\end{gathered}
\end{equation}

\subsubsection{Proportional Variation}

The third variation, P-SmvMDE, requires no additional parameters. Instead of utilizing a threshold to filter subvectors, it allocates them in subsets based on the number of samples contained in each combination that are retrieved from designated channels and applies a proportional factor to each category. With $m$ being the length of each subvector and $h$ being the number of samples extracted from designated channels, this factor is defined as $\frac{h}{m}$.

Therefore, the values of the proportional factor range from a minimum of $0$ to a maximum of $1$ and the total number of subsets in which the subvectors are allocated is equal to $m+1$. Consequently, for each $Z(j)$: $\phi_h(j) (h=1,\dots l_h)$ subvectors are formulated from each subset with $l_h$ being the length of the respective subset. Hence, the maximum value of instances ($\alpha$) for a dispersion pattern becomes $ \alpha~=~\sum_{h = 0}^{m} {(N-(m-1)d)\ (l_h \frac{h}{m})}$.

The relative frequency of each unique dispersion pattern is calculated by counting dispersion pattern instances in subvectors of each subset multiplied by their respective ($\frac{h}{m}$) factor, divided by the maximum value of instances:

\begin{equation}
\begin{gathered}
p({{\pi }_{{{v}_{0}}\dots{{v}_{m-1}}}}) = \frac{1}{\alpha} \cdot \\ \sum_{h = 0}^{m}  { \#\{j\left| j\le N-(m-1)d,\phi_h(j)\text{ has type }{{\pi }_{{{v}_{0}}\dots{{v}_{m-1}}}} \right.\} \cdot \frac{h}{m}}      
\end{gathered}
\end{equation}

\subsection{Synthetic Time-Series Experiments}

The SmvMDE variations and the original mvMDE are applied to synthetic time-series, to study the differences in their operation and their multiscale outputs. 

\subsubsection{Uncorrelated white Gaussian and $1/f$ noise}

\label{Synthetic_Experiments}

We use combinations of uncorrelated white Gaussian noise (WGN) and $1/f$ noise due to their differences in complexity and irregularity. Complexity in a time-series arises from consistent structural dynamics and therefore, when measured, is expected to follow a stable multiscale profile \cite{costa_multiscale_2002,costa_multiscale_2005}. Irregularity consists of random fluctuations that do not arise from underlying structural dynamics and is expected to have a decreasing multiscale profile. The complexity of $1/f$ noise is higher than WGN while the irregularity of WGN is higher than $1/f$ \cite{fogedby_phase_1992,ahmed_multivariate_2011}. Thus, multivariate combinations of WGN and $1/f$ time-series have been used in previous research to test multiscale entropy quantification algorithms \cite{azami_refined_2017,humeau-heurtier_multivariate_2016}. 

\subsubsection{Formulation of Experimental Setups}

To test the operation of mvMDE, all possible combinations of  WGN and $1/f$ noise are formulated in 3-channel time-series, resulting in the following inputs for the experiments:

\begin{enumerate}

\item Three WGN channels.
\item Two WGN and one $1/f$ channels.
\item One WGN and two $1/f$ channels.
\item Three $1/f$ channels. 

\end{enumerate}

Considering the operation of SmvMDE, the output entropy value will be affected to a larger degree by channels allocated to the core stratum over the periphery. This would not affect experimental setups 1) and 4). However, it would lead to different results for setups 2) and 3) which contain both WGN and $1/f$ channels based on their allocation to strata. Therefore, for SmvMDE variations, experimental setups 2) and 3) are expanded. In a first iteration, the designated channel assigned to the core is one of the WGN channels, followed by a second iteration where a $1/f$ channel is designated. This results in a total of six experimental setups for SmvMDE:

\begin{enumerate}

\item Three WGN channels.

\item Two WGN and one $1/f$ channels with WGN designated.

\item One WGN and two $1/f$ channels with WGN designated.

\item Two WGN and one $1/f$ channels with $1/f$ designated.

\item One WGN and two $1/f$ channels with $1/f$ designated.

\item Three $1/f$ channels. 

\end{enumerate}

\subsubsection{Statistical Analysis}
Each experimental setup is repeated 40 times independently and the respective mean and standard deviation are calculated for each $\tau$ value ($1$ to $20$). All experimental setups are replicated for channel lengths of 15,000 and 300 samples to assess potential differences due to long versus short time-series. The parameter values used for mvMDE and SmvMDE are chosen based on the limitations introduced by the short length time-series and match those used in the original mvMDE study to allow for easy comparison between both studies \cite{azami_multivariate_2019}. They are displayed in Table~\ref{tab_parameters}.

\begin{table}
\caption{Parameters values for Synthetic Time-Series (ST) Waveform Physiological Time-Series (WPT) and Derivative Physiological Data (DPD) Experiments.}
\label{tab_parameters}
\setlength{\tabcolsep}{8pt}
\begin{tabular}{ccccc}
\hline

Parameter (Symbol) & ST & WPT & DPD \\

\hline
Embedding Dimension ($m$) & 2 & 3 & 3 \\

Number of classes ($c$) & 5 & 6 & 4\\

Time Delay ($d$) & 1 & 1 & 1\\

Scale Factor Range ($\tau$) & 1 to 20 & 1 to 10 & 1 \\

Threshold (T and ST SmvMDE) ($t$) & 1 & 2 & 2 \\

Reduced Weight (ST SmvMDE) ($w$) & 0.5 & 0.5 & 0.5\\

\hline

\end{tabular}
\end{table}

\subsubsection{Computational Time Experiments}
To ensure that SmvMDE variations maintain the low computation time properties of the original mvMDE, 2-channel, 5-channel, and 8-channel time-series are formulated from uncorrelated WGN with channel lengths ranging from 1,000 up to 100,000 samples. Each experimental setup is repeated over 20 independent realizations and the average computation time is calculated and reported for the mvMDE and SmvMDE algorithms. For the implementation of SmvMDE algorithms, an arbitrary designated channel is selected. The computations are carried out using a PC with Intel(R) Core(TM) i7-8750H CPU @ 2.2 GHZ, 16 GB RAM running MATLAB R2018b. The parameter values of mvMDE and SmvMDE remain the same with the exception of $\tau_{max}$ being reduced from $20$ to $10$ to be consistent with \cite{azami_multivariate_2019}.

\subsection{Waveform Physiological Time-Series Experiments}

Experiments are conducted on waveform physiological time-series to study the extend to which the SmvMDE algorithms have increased discrimination capacity between physiological states. We benchmark the effect size difference of output distributions extracted using SmvMDE to those extracted using mvMDE. 

\subsubsection{MIT-BIH Polysomnographic Database}

To access multi-channel time-series formulated from high sampling rate signals recorded from different organs, the publicly available MIT-BIH Polysomnographic Database is used. It contains a total of 18 records of multiple physiological waveforms, initially recorded for the evaluation of chronic obstructive sleep apnea (OSA) syndrome and sampled at 250 Hz \cite{ichimaru_development_1999,goldberger_physiobank_2000}.

For the purpose of this study, we select the records slp41 and slp45 due to the availability of extensive sections of healthy stage 2 sleep; and the records slp04 and slp16 due to the existence of multiple incidents of OSA with arousal during stage 2 sleep. All records contain complete and synchronized recordings of EEG, ECG, BP, and RESP signals. The EEG signal is split into the frequency bands of: delta (0.5-3.5 Hz), theta (4-7.5 Hz), alpha (8-11.5 Hz), sigma (12-15.5 Hz), and beta (16-19.5 Hz) \cite{bashan_network_2012}. Hence, 8-channel time-series are extracted from each record consisting of the channels: Delta, Theta, Alpha, Sigma, Beta, ECG, BP, and RESP.

\subsubsection{Formulation and Selection of Analysis Windows}
These time-series are split into 8-channel non-overlapping windows with 7,500 samples per channel corresponding to the 30-second annotation interval of the database. Based on the annotations, we extracted 235 multi-channel ``healthy'' windows corresponding to healthy stage 2 sleep (slp41 = 96 windows, slp45 = 139 windows), and 235 multi-channel ``apnea'' windows corresponding to OSA with arousal during stage 2 sleep (slp04 = 140 windows, slp16 = 95 windows). 

\subsubsection{Calculation of DisEn}

The parameter values for the extraction of multiscale entropy distributions from the 235 ``healthy'' and 235 ``apnea''  windows are chosen based on the considerations discussed in Subsection~\ref{discuss_implem} and displayed in Table~\ref{tab_parameters} under the waveform physiological time-series (PT) column. Per window, we obtain ten values, one for each $\tau$ ($1$ to $10$).

We use mvMDE to obtain one multiscale distribution from the ``healthy'' and one from the ``apnea'' datasets. For the effective study of SmvMDE variations (T, ST, and P),
the variations are applied in eight iterations each per dataset. During each iteration a different channel is designated. This leads to the extraction of eight multiscale distributions from each dataset to study how the prioritization of each channel's dynamics affects the output entropy values and the physiological differentiation capacity of SmvMDE. 

\subsubsection{Statistical Analysis}

To effectively benchmark the differentiation capacity of SmvMDE variations to mvMDE, the following steps are completed for each $\tau$ separately:

\begin{enumerate}

\item We compute the Hedges'$g$ effect size \cite{hedges_distribution_1981} for the ``healthy'' versus ``apnea'' output distributions.

\item We calculate the effect size difference when moving from mvMDE to a certain SmvMDE variation with a particular designated channel.

\item We estimate the confidence intervals for each calculated effect size difference to verify their significance. 

\end{enumerate}

In Step 3, bootstrapping is applied to the ``healthy'' and ``apnea'' output distributions to estimate the confidence intervals. The bootstrapping is implemented by sampling with replacement the sets of 235 multiscale entropy values in each output distribution. For each output distribution of the SmvMDE variations, 40 independent realizations of bootstrapped distributions are generated. No bootstrapping is applied to the output distribution of mvMDE since we seek to benchmark the SmvMDE distributions to the same, original mvMDE results.

To implement this analysis, the bootstrapped distributions of each SmvMDE and the original distribution of mvMDE are used in the following steps, which are applied for each designated channel selection and at each $\tau$ ($1$ to $10$):

\begin{enumerate}

\item Each of the 40 bootstrapped ``healthy'' distributions is paired at random with one of the 40 bootstrapped ``apnea'' distributions. (This pairing is kept the same across all SmvMDE variations for consistency.)

\item The Hedges’ g effect size is calculated between the two distributions of each pair, resulting in 40 sets of Hedges' g effect size values.

\item Hedges'$g$ effect size values are also computed between the ``healthy'' and ``apnea'' distributions of mvMDE. 

\item The benchmarking effect size values of mvMDE are subtracted from the effect size values extracted from each pair of boostrapped distributions. This results in 40 multiscale sets of effect size differences whose mean and 95\% confidence intervals are calculated.

\end{enumerate}

We plot the mean and 95\% confidence intervals of the effect size difference separately for each designated channel selection and  $\tau$ value ($1$ to $10$).

\subsection{Derivative Physiological Data Experiments }

The operation of SmvMDE is also studied for low-temporal resolution, derivative data. The performance of SmvMDE variations is benchmarked to that of mvMDE via the difference in output entropy for separate individuals, when moving from physiological states of low to high external stress.

\subsubsection{Maximal Exercise Dataset}
For the application of SmvMDE to derivative physiological data the publicly available Treadmill Maximal Exercise Test Dataset is used \cite{moody_physionet/computing_2010,mongin_heart_2021}. This dataset was collected, curated, and published by the Exercise Physiology and Human Performance Lab of the University of Malaga. The recordings include five cardiorespiratory variables: heart rate (HR, in beats per min), oxygen consumption (VO2, in mL/min), carbon dioxide production (VCO2, in mL/min), respiration rate (RR, in respirations/min), and pulmonary ventilation (VE, in L/min). All variables were recorded in a synchronized manner with the sampling event being each breath measurement, resulting in a varied sampling period (usually in the range of 1-4 s).

Each test consisted of an individual walking and running on a treadmill, starting with a warm-up period of treadmill speeds close to 5 km/h, followed by a period of gradual speed increase that reached speeds in the range of 14 to 17 km/h, and completed with a cool-down period with speeds close to 5km/h. A total of 857 individuals participated in the study with some people having more than one test, resulting in 992 recordings. The participants' ages ranged from 10 to 63 y.o.

\subsubsection{Formulation and Selection of Analysis Windows}

Two physiological state classes are formulated: a low speed (LS) class that corresponds to data recorded during warm-up until the speed reached 7 km/h; and a high speed (HS) class that corresponds to data recorded while the treadmill speed was higher than 15 km/h. We selected recordings with at least 120 synchronised samples for each class to ensure an adequate window size for analysis. In the few cases where an individual had more than one eligible test, the first one was selected. A total of 98 eligible recordings with age range 14 to 50 y.o.~are extracted. 

\subsubsection{Calculation of DisEn}
The extracted data are 98 pairs of multivariate 120-sample windows, with each pair including one segment from the LS class and one from the HS. Due to the low temporal resolution of the data and the consequent small window size, the analysis is done only at temporal scale $\tau=1$. The selected parameter values are displayed in Table~\ref{tab_parameters}. All SmvMDE variations (T, ST, and P) are applied in five iterations each, during which a different channel is designated. Consequently, for each algorithm and designated channel selection 98 pairs of DisEn values are extracted.

\subsubsection{Statistical Analysis}

For each experimental setup and within each of the 98 pairs of DisEn values, the entropy difference observed when moving from the LS to the HS state is recorded. Boxplots are generated to compare the output difference distributions between SmvMDE and mvMDE for each designated channel selection. Additionally, the mean absolute difference observed in each difference distribution and the number of entries that displayed an increased absolute value of difference during each SmvMDE configuration, compared to their mvMDE values, are reported. Finally, to highlight a potential directionality that could match the EP hypothesis \cite{fossion_physicists_2018,barajas-martinez_physiological_2022}, the number of entries with a higher entropy value in the LS state than in the HS state are also reported for each configuration.

\section{Results and Discussion}

\subsection{Synthetic Time-Series Experiments}
\label{Synthetic_results}
\begin{figure*}[!t]
\centerline{\includegraphics[width=2\columnwidth]{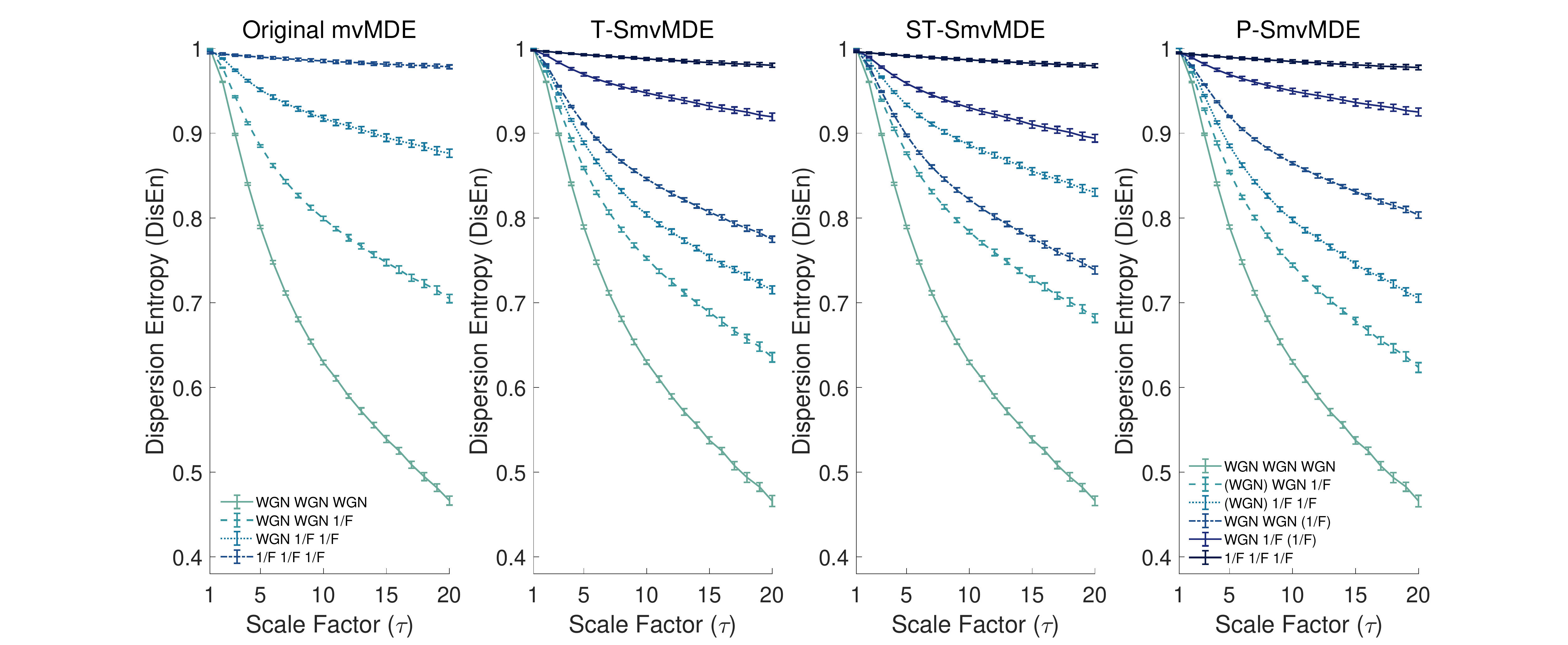}}
\caption{The mean and standard deviation of output DisEn are plotted for $\tau$ values of $1$ to $20$ for the four experimental setups of mvMDE and the six experimental setups of SmvMDE with time-series length of 15,000 samples. For SmvMDE, the designated channel is displayed within (). }
\label{Synthetic_15K}
\end{figure*}

The results of the application of mvMDE and SmvMDE on 3-channel time-series of WGN and $1/f$ noise, are presented in Fig.~\ref{Synthetic_15K} and Fig.~\ref{Synthetic_300} for univariate length of 15,000 and 300 samples, respectively. For each experimental setup, replicated for 40 independent iterations, the mean and standard deviation of DisEn values are plotted for each $\tau$ ($1$ to $20$).

\subsubsection{mvMDE Operation}

The operation of the mvMDE matches the patterns that have been verified by prior research \cite{azami_multivariate_2019}. As $\tau$ increases, the output entropy value has a stronger decline for the 3-channel WGN time-series. As the number of $1/f$ channels increase, the output entropy value follows a more stable profile with the 3-channel $1/f$ time-series being the most stable.

\subsubsection{SmvMDE Operation}

For experimental setups that contain solely WGN channels and $1/f$ channels respectively, the operation of all three SmvMDE variations is identical to mvMDE, as expected. In contrast in the other experiments, a stronger decline of output entropy is observed as $\tau$ increases when a WGN channel is designated. Instead when a $1/f$ channel is designated, the output follows a more stable profile for increasing values of $\tau$.

When comparing the results of the three SmvMDE variations for the same experimental setup:

\begin{enumerate}

\item Using the mvMDE output values as reference, the largest deviations are observed by the P-SmvMDE variation, followed by the T-SmvMDE, and then the ST-SmvMDE variation.

\item The ST-SmvMDE outputs are between those of T-SmvMDE and mvMDE as expected by its design and the $w$ value set to $0.5$.

\item The higher deviation of the P-SmvMDE outputs from T-SmvMDE is expected when considering that for an $m = 2$ the P-SmvMDE variation gives a higher prioritization to the core stratum than the respective implementation of T-SmvMDE with $m =2$ and $t = 1$.
\end{enumerate}

\subsubsection{Short Length Time-Series}
\begin{figure*}[!t]
\centerline{\includegraphics[width=2\columnwidth]{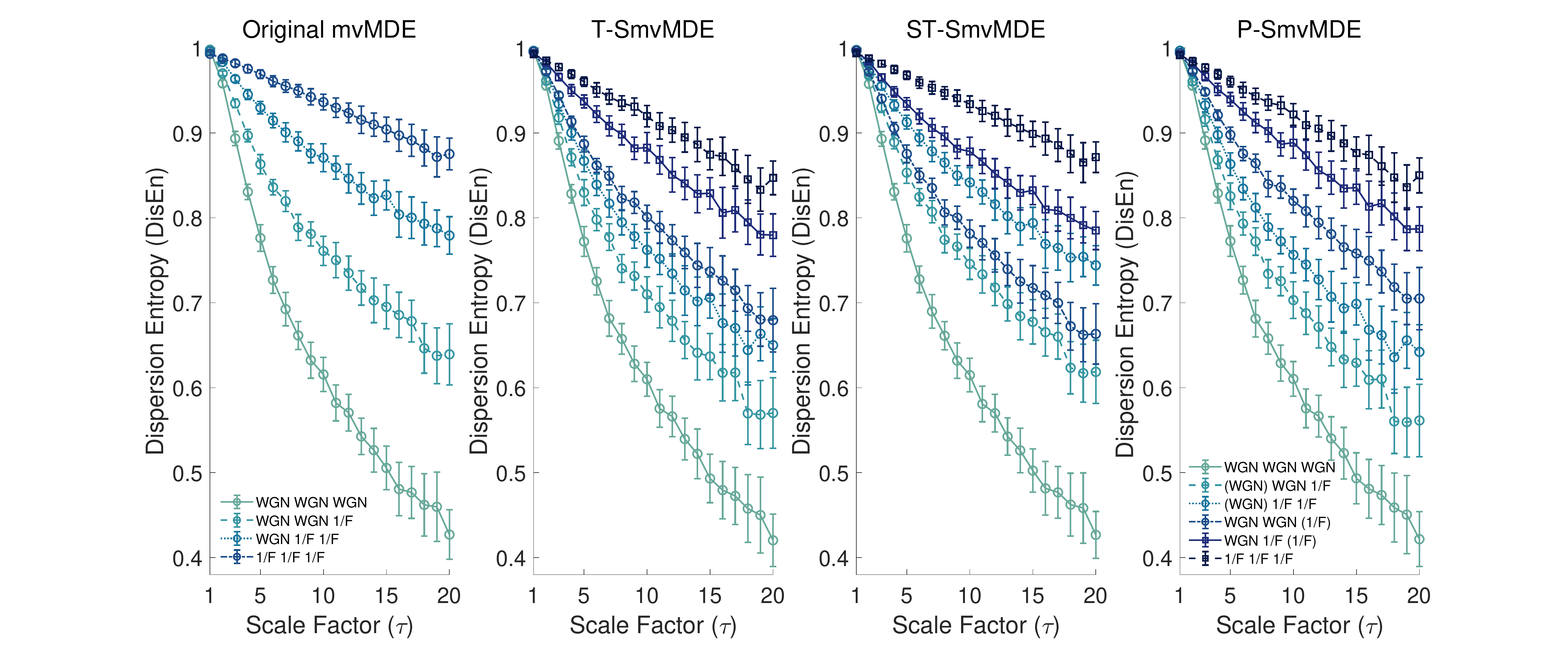}}
\caption{The mean and standard deviation of output DisEn are plotted for $\tau$ values of $1$ to $20$ for the four experimental setups of mvMDE and the six experimental setups of SmvMDE with time-series length of 300 samples. For SmvMDE, the designated channel is displayed within (). }
\label{Synthetic_300}
\end{figure*}

Fig.~\ref{Synthetic_300} displays the results for the 300 sample length experiments. For all tested algorithms, the outputs follow the same patterns as their 15,000 sample length equivalent, indicating that the operation of SmvMDE remains the same regardless of time-series length. However, for all experimental setups, the standard deviation values are increased, with the increase being stronger for larger $\tau$ values, as expected. Consequently, between the outputs of SmvMDE variations, overlapping can be observed between experimental setups that combine WGN and $1/f$. This indicates that during the analysis of multi-channel time-series, the sample size of the window being analyzed should be larger than the respective minimum size for mvMDE.

\subsection{Computational Time}

\begin{table}
\caption{Computational time of mvMDE and SmvMDE in seconds.}
\label{CT_Results}
\setlength{\tabcolsep}{4.6pt}
\begin{tabular}{ccccc}
\hline
Samples and Channels & mvMDE & T & ST & P \\

\hline

1,000 samples and 2-channels & 
0.024 & 0.025 & 0.026 & 0.026\\

1,000 samples and 5-channels &
0.061 & 0.059 & 0.064 & 0.065\\

1,000 samples and 8-channels &
0.100 & 0.0095 & 0.117 & 0.126\\
\hline

3,000 samples and 2-channels &
0.067 & 0.069 & 0.069 & 0.070 \\

3,000 samples and 5-channels &
0.177 & 0.171 & 0.183 & 0.187 \\

3,000 samples and 8-channels &
0.302 & 0.285 & 0.339 & 0.356 \\
\hline

10,000 samples and 2-channels &
0.241 & 0.251 & 0.250 & 0.255 \\

10,000 samples and 5-channels &
0.649 & 0.631 & 0.663 & 0.685 \\

10,000 samples and 8-channels &
1.182 & 1.029 & 1.240 & 1.254 \\
\hline

30,000 samples and 2-channels &
1.056 & 1.091 & 1.101 & 1.108  \\

30,000 samples and 5-channels &
2.839 & 2.612 & 2.879 & 2.870  \\

30,000 samples and 8-channels &
4.789 & 4.526 & 5.044 & 5.044 \\

\hline

100,000 samples and 2-channels &
8.048 & 8.067 & 8.202 & 8.181 \\

100,000 samples and 5-channels &
20.550 & 20.157 & 20.820 & 20.967 \\

100,000 samples and 8-channels &
33.571 & 32.436 & 35.243 & 35.133 \\

\hline

\end{tabular}
\label{tab3}
\end{table}

The results in Table~\ref{CT_Results} indicate that SmvMDE variations maintain the low computational time of the original mvMDE, as expected, since no computationally critical operations have been modified and the linear time complexity is maintained. Across all variations the main factor affecting the computation time is the univariate length of the time-series. When comparing the results for experimental setups with the same univariate length, the differences in computation time between the original mvMDE and the SmvMDE variations become more noticeable for higher number of channels.

The maximum differences in computation time are noted in the experimental setup with a time-series length of 100,000 samples and 8-channels. The maximum increase of $1.672$ seconds ($4.98\%$) is noted when moving from the mvMDE to the ST-SmvMDE algorithm while the maximum decrease of $1.135$ seconds ($3.38\%$) is noted when moving from mvMDE to T-SmvMDE. The decrease of computation time in the case of T-SmvMDE is an expected benefit due to the lower number of subvectors utilised in that variation. 

\subsection{Waveform Physiological Time-Series Experiments}
\label{HRPT}

\begin{figure*}[!t]
\centerline{\includegraphics[width=2.2\columnwidth]{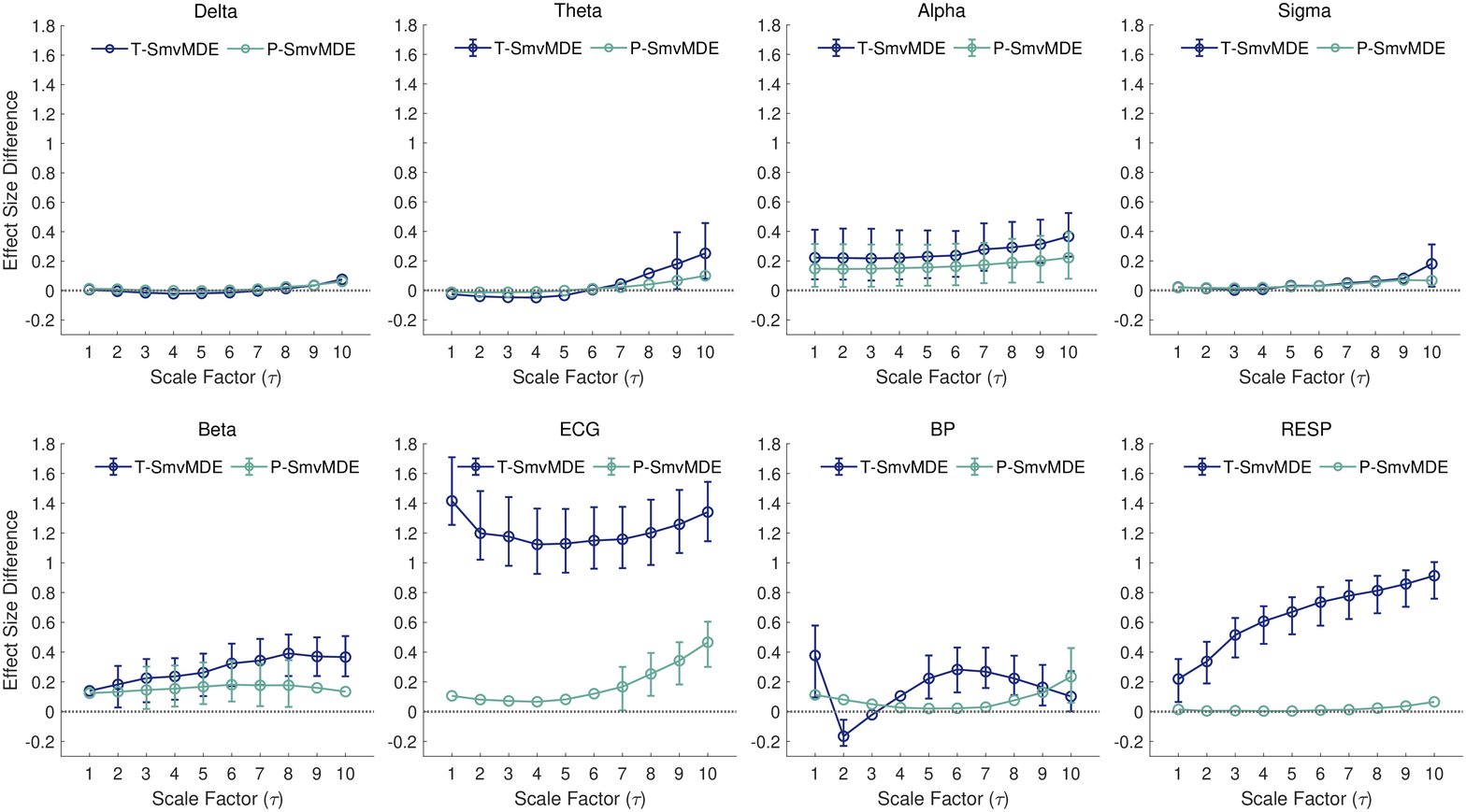}}
\caption{The mean and the 95\% confidence intervals of the effect size difference calculated from subtracting the multiscale mvMDE Hedges' $g$ effect sizes from those of T-SmvMDE and P-SmvMDE. Each subplot corresponds to a different designated channel selection.}
\label{EffectSize_Diff_PT}
\end{figure*}

The results of the statistical analysis implemented on the output entropy distributions extracted from the 235 ``healthy'' and 235 ``apnea'' 8-channel windows using T-SmvMDE and P-SmvMDE, are presented in Fig.~\ref{EffectSize_Diff_PT}, with each subplot corresponding to a different designated channel selection. The mean Hedges'$g$ effect size difference and the $95\%$ confidence intervals are plotted for each $\tau$ ($1$ to $10$).For clarity, only the confidence intervals that do not overlap with $0$ are plotted.

ST-SmvMDE is, by design, an intermediary variation between T-SmvMDE and mvMDE. Thus, its outputs also follow an intermediary pattern, closer to the operation of mvMDE, leading to smaller effect size differences.

\subsubsection{T-SmvMDE Operation}
The benchmarking of T-SmvMDE, indicates that the prioritization of the following channels leads to consistent increases in differences between the output entropy distributions extracted from "healthy" vs "apnea" windows when moving from the application of mvMDE to T-SmvMDE:

\begin{itemize}
    \item ECG, RESP, and Alpha channels across all values of $\tau$.
    \item Beta channel for $\tau$ values of $2$ to $10$.
    \item BP channel for $\tau$ values of $5$ to $10$.
\end{itemize}

The multiple cases of increase in effect size indicate that the T-SmvMDE variation may quantify differences between the two states that the direct application of mvMDE was not able to highlight.

\subsubsection{P-SmvMDE Operation}
The respective benchmarking results for P-SmvMDE indicate that increases in difference between the output entropy distributions are observed when prioritizing one of the following channels:

\begin{itemize}
    \item Alpha across all values of $\tau$.
    \item Beta for $\tau$ values of 3 to 8.
    \item ECG for $\tau$ values of 7 to 10.
\end{itemize}

Consequently, the designated channels displaying increased discrimination capacity for P-SmvMDE were also highlighted by T-SmvMDE. However, increases observed by moving to P-SmvMDE were smaller in magnitude and for fewer designated channel selections compared to T-SmvMDE. Considering the parameter values used for the SmvMDE variations in this setup, the T-SmvMDE sets a higher prioritization to the core stratum over the periphery compared to P-SmvMDE. This may indicate that this particular application benefited from implementations that defined stronger prioritization. Furthermore, within the framework of Stratified Entropy, the detection of certain prioritization cases as more effective in extracting distinct feature distributions between physiological states, highlights the potential for the development of feature selection methodologies that would aim to optimize physiological classification tasks.

\subsection{Derivative Physiological Data Experiments}
\label{LRPT}

\begin{figure*}[!t]
\centerline{\includegraphics[width=2.0\columnwidth]{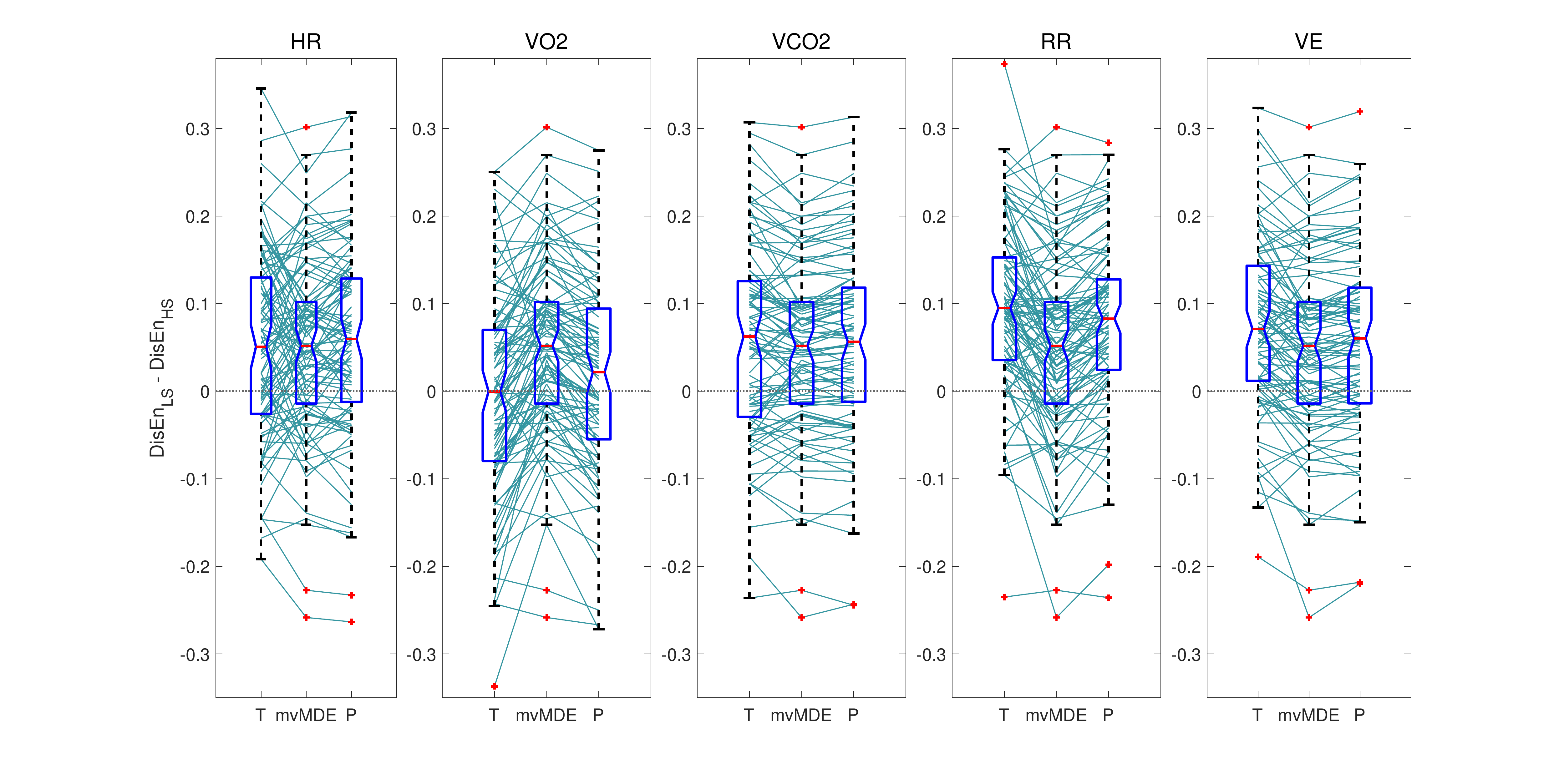}}
\caption{Boxplots comparing the DisEn Entropy difference calculated when moving from LS to HS using T-SmvMDE, mvMDE and P-SmvMDE. The plots correspond to different designated channel selections with the benchmarking values of mvMDE remaining the same. The lines connect the difference of the same experimental pair across the output differences of each algorithm.}
\label{LSHS_Boxplot}
\end{figure*}

The DisEn differences observed when moving from the LS to the HS state for each of the 98 exercise tests are displayed in the boxplots of Fig.~\ref{LSHS_Boxplot}. Each panel corresponds to a different designated channel selection and includes the boxplots with the distributions of differences observed through the application of mvMDE, T-SmvMDE, and P-SmvMDE. The mean absolute difference observed during the application of mvMDE is equal to $0.0894$ while the respective mean absolute differences, number of entries with increased entropy difference compared to mvMDE and number of entries with larger entropy during LS versus HS are displayed for each SmvMDE and designated channel in Table~\ref{dpd_parameters}. 

\begin{table}
\caption{Mean absolute difference, Number of Entries with increased entropy difference compared to mvMDE, and Number of entries with larger entropy during LS versus HS.}
\label{dpd_parameters}
\setlength{\tabcolsep}{8pt}
\begin{tabular}{cccc}
\hline

T-SmvMDE & Mean Absolute Difference & Improved & Positive \\

\hline
HR & $0.093$ & $46$ & $64$ \\

VO2 & $0.093$ & $51$ & $49$ \\

VCO2 & $0.100$ & $68$ & $64$\\

RR & $0.115$ & $66$ & $87$ \\

VE &  $0.097$ & $54$ & $78$ \\

\hline

P-SmvMDE & Mean Absolute Difference & Improved & Positive \\

\hline

HR & $0.097 $ & $57$ & $71$\\

VO2 & $0.086$ & $40$ & $51$\\

VCO2 & $0.095$ & $69$ & $67$\\

RR & $0.103$ & $64$ &  $81$\\

VE &  $0.093$ & $53$ & $69$\\

\hline

\end{tabular}
\end{table}

\subsubsection{SmvMDE Operation}

When benchmarking the operation of SmvMDE to mvMDE, an improvement in differentiation capacity is noted when the VCO2, RR, and VE channels are designated. This improvement is consistent for both T-SmvMDE and P-SmvMDE with increases in the mean absolute difference and the entries with increased LS-HS difference. The selection of HR as a designated channel displayed increased differentiation capacity for P-SmvMDE.


Similarly to Subsection~\ref{HRPT}, the majority of designated channels for which an increase in the discrimination capacity of SmvMDE is noted are common between T-SmvMDE and P-SmvMDE, indicating that while the two variations provide different ways to prioritise strata, they have the capacity of highlighting similar dynamics that were overshadowed by traditional multivariate analysis.

It is important to note that when designating the RR channel, the largest mean absolute difference is observed for both T-SmvMDE and P-SmvMDE as well as the largest number of entries where the LS DisEn values are higher than the HS ones. This points towards a LoC process \cite{lipsitz_dynamics_2002,goldberger_what_2002} when moving from a steady state to a state that induces increased stress in the system, in alignment with the EP hypothesis \cite{fossion_physicists_2018,barajas-martinez_physiological_2022}.

\subsection{On the implementation of Stratified Entropy}
\label{discuss_implem} 

\subsubsection{Input Window Length}

For the implementation of entropy quantification, the selection of the $c$ and $m$ parameters defines the minimum length of each channel within the input window. The univariate DisEn algorithm is capable of analysing short-length time-series \cite{azami_amplitude-_2018} with minimum length ($L$) being $L > c^m \cdot \tau_{max}$. The mvMDE variation of the algorithm further improved its capacity to operate on short-length time-series due to the utilization of larger-multivariate embedding vectors compared to their univariate counterparts \cite{azami_multivariate_2019} leading to a minimum length of: $L > \frac{c^m \cdot \tau_{max}}{\binom {m\cdot p}{m}}$. For SmvMDE variations, the minimum input length is between the limits of univariate DisEn and mvMDE. As shown in Subsection~\ref{Synthetic_results}, overlapping is observed in the short-length time-series among the large $\tau$ value outputs while analyzing the same time-series with different channels being prioritized. Consequently, we would recommend the utilization of a stricter minimum, closer to the univariate DisEn: $L > c^m \cdot \tau_{max}$ when deploying SmvMDE variations.

\subsubsection{Number of designated channels}

With $m$ being an exponent in defining the minimum input window length, it is important to consider that during the implementation of Stratified Entropy, an increased value of $m$ might be needed when increasing the number of designated channels. In such case, it is important ensure that there are not multiple subvectors consisting entirely of samples retrieved from designated channels which would lead to them being treated equally and result to an output profile that would resemble that of mvMDE. Furthermore, while an increase in the value of $m$ would allow additional designated channels this might not be an optimal approach, since an overshadowing of dynamics would now be possible to occur within the core stratum itself. Thus, we recommend that the majority of Stratified Entropy applications follow a conservative approach when allocating channels to the core stratum.

\subsubsection{Number of Strata}
\label{Strata_Number}
The total number of strata defined in a Stratified Entropy implementation affects both its design and its implementation since appropriate algorithmic steps have to be formulated for the prioritization of channels based on their strata allocation, while proper selection of parameter values is required for effective operation. For example, in the case of expanding the presented SmvMDE variations from a two to a three strata configuration, the T-SmvMDE and the ST-SmvMDE variations could be modified to operate with two different $t$ and $w$ (in the case of ST) values based on which strata are prioritized, while P-SmvMDE could be modified with having two tiers of proportional factors respectively. This design modification could be complemented with an appropriate increase of the $m$ value to allow samples of varied prioritization to be included in the same subvectors similarly to the process discussed for having multiple designated channels.

However, while the expansion of the total number of strata is possible, it increases algorithmic complexity and restricts the range of effective parameter values, particularly of $m$ as discussed above. Therefore, configurations with increased numbers of strata might be more relevant for applications that would clearly benefit from their utilization despite the increased complexity, such as for example when a priori knowledge exists with regards to a hierarchy of channels.

\subsection{Limitations and Future Work}

Our algorithmic variations illustrate successful implementations of the Stratified Entropy framework, with effective prioritization of the channels allocated to the core stratum over the periphery, and the extraction of novel features. However, it is important to expand its implementation using additional entropy quantification algorithms, such as PEn, to acquire a more complete perspective on the utility that the framework offers. Furthermore, due to its capacity to be applied in a modular manner and with low computational cost, it would be worthwhile to combine Stratified Entropy with other variations of entropy algorithms to target specific applications. Examples include its integration with the aforementioned non-uniform multiscale embedding to incorporate a priori knowledge, with optimal scale selection for each channel, or the utilization of a fuzzy membership function in DisEn \cite{rostaghi_fuzzy_2021}.

The results in the experiments of derivative physiological data indicate a directionality in agreement with the LoC paradigm and the EP hypothesis. Hence, it would be important to replicate the analyses in other datasets and study the capacity of SmvMDE to quantify the directionality of EP phenomena. Moreover, the combination of SmvMDE with machine learning would allow for physiological state classification and prediction tasks. Consequently, strata allocations should be selected with appropriate justification or through effective feature selection processes to avoid data dredging or overfitting. 

\section{Conclusion}

We introduce the framework of Stratified Entropy and present three algorithmic variations for its implementation. Stratified Entropy allows the extraction of features that would not be accessible through traditional multivariate entropy analysis by allowing the prioritization of certain channels' dynamics over others' based on the allocation of channels to different strata. The SmvMDE variations significantly extend mvMDE through the inclusion of algorithmic steps that prioritize samples extracted from channels in the prioritized stratum during the calculation of the entropy value.

The results from the application of SmvMDE to time-series consisting of uncorrelated WGN and $1/f$ noise indicate that the variations successfully prioritize the dynamics of the designated channel. The low computation time profile of the original mvMDE variation is maintained due to no computationally critical steps being modified. When applying the SmvMDE variations to 8-channel waveform physiological time-series, certain SmvMDE features produce distributions with higher statistical difference between healthy versus OSA sleep of stage 2, indicating increased discrimination capacity of SmvMDE over mvMDE for applications that would benefit from a stratification of the time-series' channels. The respective results from low temporal resolution derivative physiological data further highlight the increased discrimination capacity of SmvMDE and its potential use to detect the directionality of ``entropy pump phenomena''. 

The presented framework is flexible with regards to the number of channels allocated to the prioritized stratum and the total number of strata. Furthermore, it can be extended to other entropy quantification algorithms and combined with machine learning. Consequently, with appropriate algorithmic design and parameter configuration, we expect the framework of Stratified Entropy to provide novel and effective methodologies for the extraction of viable physiological information. 

\section*{Acknowledgment}

We thank John Stewart Fabila Carrasco and Ian Piper for discussions on mathematical notation and  physiological signals, respectively, and to the Reviewers for their feedback.



\bibliographystyle{IEEEtran}
\bibliography{references_nou}

\begin{thebibliography}{10}
\providecommand{\url}[1]{#1}
\csname url@samestyle\endcsname
\providecommand{\newblock}{\relax}
\providecommand{\bibinfo}[2]{#2}
\providecommand{\BIBentrySTDinterwordspacing}{\spaceskip=0pt\relax}
\providecommand{\BIBentryALTinterwordstretchfactor}{4}
\providecommand{\BIBentryALTinterwordspacing}{\spaceskip=\fontdimen2\font plus
\BIBentryALTinterwordstretchfactor\fontdimen3\font minus
  \fontdimen4\font\relax}
\providecommand{\BIBforeignlanguage}[2]{{%
\expandafter\ifx\csname l@#1\endcsname\relax
\typeout{** WARNING: IEEEtran.bst: No hyphenation pattern has been}%
\typeout{** loaded for the language `#1'. Using the pattern for}%
\typeout{** the default language instead.}%
\else
\language=\csname l@#1\endcsname
\fi
#2}}
\providecommand{\BIBdecl}{\relax}
\BIBdecl

\bibitem{yetisen_wearables_2018}
A.~K. Yetisen, J.~L. Martinez-Hurtado, B.~Ünal, A.~Khademhosseini, and
  H.~Butt, ``\BIBforeignlanguage{en}{Wearables in {Medicine}},''
  \emph{\BIBforeignlanguage{en}{Adv. Mater.}}, vol.~30, no.~33, p. 1706910,
  Aug. 2018.

\bibitem{witt_windows_2019}
D.~R. Witt, R.~A. Kellogg, M.~P. Snyder, and J.~Dunn,
  ``\BIBforeignlanguage{en}{Windows into human health through wearables data
  analytics},'' \emph{\BIBforeignlanguage{en}{Current Opinion in Biomedical
  Engineering}}, vol.~9, pp. 28--46, Mar. 2019.

\bibitem{paine_systematic_2016}
C.~W. Paine, V.~V. Goel, E.~Ely, C.~D. Stave, S.~Stemler, M.~Zander, and C.~P.
  Bonafide, ``\BIBforeignlanguage{en}{Systematic {Review} of {Physiologic}
  {Monitor} {Alarm} {Characteristics} and {Pragmatic} {Interventions} to
  {Reduce} {Alarm} {Frequency}: {Review} of {Physiologic} {Monitor}
  {Alarms}},'' \emph{\BIBforeignlanguage{en}{Journal of Hospital Medicine}},
  vol.~11, no.~2, pp. 136--144, Feb. 2016.

\bibitem{cerutti_multivariate_2012}
S.~Cerutti, ``\BIBforeignlanguage{en}{Multivariate and multiscale analysis of
  biomedical signals: {Towards} a comprehensive approach to medical
  diagnosis},'' in \emph{\BIBforeignlanguage{en}{2012 25th {IEEE}
  {International} {Symposium} on {Computer}-{Based} {Medical} {Systems}
  ({CBMS})}}.\hskip 1em plus 0.5em minus 0.4em\relax Rome, Italy: IEEE, Jun.
  2012, pp. 1--5.

\bibitem{pereda_nonlinear_2005}
E.~Pereda, R.~Q. Quiroga, and J.~Bhattacharya,
  ``\BIBforeignlanguage{en}{Nonlinear multivariate analysis of
  neurophysiological signals},'' \emph{\BIBforeignlanguage{en}{Progress in
  Neurobiology}}, vol.~77, no. 1-2, pp. 1--37, Sep. 2005.

\bibitem{cerutti_multiscale_2009}
S.~Cerutti, D.~Hoyer, and A.~Voss, ``\BIBforeignlanguage{en}{Multiscale,
  multiorgan and multivariate complexity analyses of cardiovascular
  regulation},'' \emph{\BIBforeignlanguage{en}{Phil. Trans. R. Soc. A.}}, vol.
  367, no. 1892, pp. 1337--1358, Apr. 2009.

\bibitem{bartsch_network_2015}
R.~P. Bartsch, K.~K.~L. Liu, A.~Bashan, and P.~C. Ivanov,
  ``\BIBforeignlanguage{en}{Network {Physiology}: {How} {Organ} {Systems}
  {Dynamically} {Interact}},'' \emph{\BIBforeignlanguage{en}{PLoS ONE}},
  vol.~10, no.~11, p. e0142143, Nov. 2015.

\bibitem{fossion_physicists_2018}
R.~Fossion, A.~L. Rivera, and B.~Estañol, ``\BIBforeignlanguage{en}{A
  physicist’s view of homeostasis: how time series of continuous monitoring
  reflect the function of physiological variables in regulatory mechanisms},''
  \emph{\BIBforeignlanguage{en}{Physiol. Meas.}}, vol.~39, no.~8, p. 084007,
  Aug. 2018.

\bibitem{goldberger_fractal_2002}
A.~L. Goldberger, L.~A.~N. Amaral, J.~M. Hausdorff, P.~C. Ivanov, C.-K. Peng,
  and H.~E. Stanley, ``\BIBforeignlanguage{en}{Fractal dynamics in physiology:
  {Alterations} with disease and aging},''
  \emph{\BIBforeignlanguage{en}{Proceedings of the National Academy of
  Sciences}}, vol.~99, no. Supplement 1, pp. 2466--2472, Feb. 2002.

\bibitem{goldberger_what_2002}
A.~L. Goldberger, C.-K. Peng, and L.~A. Lipsitz, ``\BIBforeignlanguage{en}{What
  is physiologic complexity and how does it change with aging and disease?}''
  \emph{\BIBforeignlanguage{en}{Neurobiology of Aging}}, vol.~23, no.~1, pp.
  23--26, Jan. 2002.

\bibitem{scafetta_fractal_2007}
N.~Scafetta, R.~E. Moon, and B.~J. West, ``\BIBforeignlanguage{en}{Fractal
  response of physiological signals to stress conditions, environmental
  changes, and neurodegenerative diseases},''
  \emph{\BIBforeignlanguage{en}{Complexity}}, vol.~12, no.~5, pp. 12--17, May
  2007.

\bibitem{martis_application_2011}
R.~J. Martis, U.~R. Acharya, A.~K. Ray, and C.~Chakraborty,
  ``\BIBforeignlanguage{en}{Application of higher order cumulants to {ECG}
  signals for the cardiac health diagnosis},'' in
  \emph{\BIBforeignlanguage{en}{2011 {Annual} {International} {Conference} of
  the {IEEE} {Engineering} in {Medicine} and {Biology} {Society}}}.\hskip 1em
  plus 0.5em minus 0.4em\relax Boston, MA: IEEE, Aug. 2011, pp. 1697--1700.

\bibitem{pradhan_higher-order_2012}
C.~Pradhan, S.~K. Jena, S.~R. Nadar, and N.~Pradhan,
  ``\BIBforeignlanguage{en}{Higher-{Order} {Spectrum} in {Understanding}
  {Nonlinearity} in {EEG} {Rhythms}},''
  \emph{\BIBforeignlanguage{en}{Computational and Mathematical Methods in
  Medicine}}, vol. 2012, pp. 1--8, 2012.

\bibitem{muller_causality_2016}
A.~Müller, J.~F. Kraemer, T.~Penzel, H.~Bonnemeier, J.~Kurths, and N.~Wessel,
  ``\BIBforeignlanguage{en}{Causality in physiological signals},''
  \emph{\BIBforeignlanguage{en}{Physiol. Meas.}}, vol.~37, no.~5, pp. R46--R72,
  May 2016.

\bibitem{azimi_missing_2019}
I.~Azimi, T.~Pahikkala, A.~M. Rahmani, H.~Niela-Vilén, A.~Axelin, and
  P.~Liljeberg, ``\BIBforeignlanguage{en}{Missing data resilient
  decision-making for healthcare {IoT} through personalization: {A} case study
  on maternal health},'' \emph{\BIBforeignlanguage{en}{Future Generation
  Computer Systems}}, vol.~96, pp. 297--308, Jul. 2019.

\bibitem{kumar_automated_2016}
R.~B. Kumar, N.~D. Goren, D.~E. Stark, D.~P. Wall, and C.~A. Longhurst,
  ``\BIBforeignlanguage{en}{Automated integration of continuous glucose monitor
  data in the electronic health record using consumer technology},''
  \emph{\BIBforeignlanguage{en}{J Am Med Inform Assoc}}, vol.~23, no.~3, pp.
  532--537, May 2016.

\bibitem{moody_physionet/computing_2010}
G.~B. Moody, ``\BIBforeignlanguage{en}{The {PhysioNet}/{Computing} in
  {Cardiology} {Challenge} 2010: {Mind} the {Gap}},''
  \emph{\BIBforeignlanguage{en}{Comput Cardiol}}, p.~13, 2010.

\bibitem{ribeiro_entropy_2021}
M.~Ribeiro, T.~Henriques, L.~Castro, A.~Souto, L.~Antunes, C.~Costa-Santos, and
  A.~Teixeira, ``\BIBforeignlanguage{en}{The {Entropy} {Universe}},''
  \emph{\BIBforeignlanguage{en}{Entropy}}, vol.~23, no.~2, p. 222, Feb. 2021.

\bibitem{shannon_mathematical_1948}
C.~E. Shannon, ``\BIBforeignlanguage{en}{A {Mathematical} {Theory} of
  {Communication}},'' \emph{\BIBforeignlanguage{en}{Bell Syst. Tech. J.}}, vol.
  vol. 27, pp. 623--656, 1948.

\bibitem{papoulis_probability_2002}
A.~Papoulis and S.~U. Pillai, \emph{\BIBforeignlanguage{en}{Probability, random
  variables, and stochastic processes}}, 4th~ed.\hskip 1em plus 0.5em minus
  0.4em\relax Boston: McGraw-Hill, 2002.

\bibitem{rostaghi_application_2019}
M.~Rostaghi, M.~R. Ashory, and H.~Azami, ``\BIBforeignlanguage{en}{Application
  of dispersion entropy to status characterization of rotary machines},''
  \emph{\BIBforeignlanguage{en}{Journal of Sound and Vibration}}, vol. 438, pp.
  291--308, Jan. 2019.

\bibitem{huo_entropy_2020}
Z.~Huo, M.~Martinez-Garcia, Y.~Zhang, R.~Yan, and L.~Shu,
  ``\BIBforeignlanguage{en}{Entropy {Measures} in {Machine} {Fault}
  {Diagnosis}: {Insights} and {Applications}},''
  \emph{\BIBforeignlanguage{en}{IEEE Trans. Instrum. Meas.}}, vol.~69, no.~6,
  pp. 2607--2620, Jun. 2020.

\bibitem{wu_modified_2018}
Y.~Wu, P.~Shang, and Y.~Li, ``\BIBforeignlanguage{en}{Modified generalized
  multiscale sample entropy and surrogate data analysis for financial time
  series},'' \emph{\BIBforeignlanguage{en}{Nonlinear Dyn}}, vol.~92, no.~3, pp.
  1335--1350, May 2018.

\bibitem{zhang_multivariate_2019}
Y.~Zhang, P.~Shang, and H.~Xiong, ``\BIBforeignlanguage{en}{Multivariate
  generalized information entropy of financial time series},''
  \emph{\BIBforeignlanguage{en}{Physica A: Statistical Mechanics and its
  Applications}}, vol. 525, pp. 1212--1223, Jul. 2019.

\bibitem{scheffer_early-warning_2009}
M.~Scheffer, J.~Bascompte, W.~A. Brock, V.~Brovkin, S.~R. Carpenter, V.~Dakos,
  H.~Held, E.~H. van Nes, M.~Rietkerk, and G.~Sugihara,
  ``\BIBforeignlanguage{en}{Early-warning signals for critical transitions},''
  \emph{\BIBforeignlanguage{en}{Nature}}, vol. 461, no. 7260, pp. 53--59, Sep.
  2009.

\bibitem{olde_rikkert_slowing_2016}
M.~G.~M. Olde~Rikkert, V.~Dakos, T.~G. Buchman, R.~d. Boer, L.~Glass, A.~O.~J.
  Cramer, S.~Levin, E.~van Nes, G.~Sugihara, M.~D. Ferrari, E.~A. Tolner,
  I.~van~de Leemput, J.~Lagro, R.~Melis, and M.~Scheffer,
  ``\BIBforeignlanguage{en}{Slowing {Down} of {Recovery} as {Generic} {Risk}
  {Marker} for {Acute} {Severity} {Transitions} in {Chronic} {Diseases}:},''
  \emph{\BIBforeignlanguage{en}{Critical Care Medicine}}, vol.~44, no.~3, pp.
  601--606, Mar. 2016.

\bibitem{nakazato_estimation_2020}
Y.~Nakazato, T.~Sugiyama, R.~Ohno, H.~Shimoyama, D.~L. Leung, A.~A. Cohen,
  R.~Kurane, S.~Hirose, A.~Watanabe, and H.~Shimoyama,
  ``\BIBforeignlanguage{en}{Estimation of homeostatic dysregulation and frailty
  using biomarker variability: a principal component analysis of hemodialysis
  patients},'' \emph{\BIBforeignlanguage{en}{Sci Rep}}, vol.~10, no.~1, p.
  10314, Dec. 2020.

\bibitem{lipsitz_dynamics_2002}
L.~A. Lipsitz, ``\BIBforeignlanguage{en}{Dynamics of {Stability}: {The}
  {Physiologic} {Basis} of {Functional} {Health} and {Frailty}},''
  \emph{\BIBforeignlanguage{en}{The Journals of Gerontology Series A:
  Biological Sciences and Medical Sciences}}, vol.~57, no.~3, pp. B115--B125,
  Mar. 2002.

\bibitem{barajas-martinez_physiological_2022}
A.~Barajas-Martínez, R.~Mehta, E.~Ibarra-Coronado, R.~Fossion, V.~J.
  Martínez~Garcés, M.~R. Arellano, I.~A. González~Alvarez, Y.~V.~M.
  Bautista, O.~Y. Bello-Chavolla, N.~R. Pedraza, B.~R. Encinas, C.~I.~P.
  Carrión, M.~I.~J. Ávila, J.~C. Valladares-García, P.~E. Vanegas-Cedillo,
  D.~H. Juárez, A.~Vargas-Vázquez, N.~E. Antonio-Villa, P.~Almeda-Valdes,
  O.~Resendis-Antonio, M.~Hiriart, A.~Frank, C.~A. Aguilar-Salinas, and A.~L.
  Rivera, ``\BIBforeignlanguage{en}{Physiological {Network} {Is} {Disrupted} in
  {Severe} {COVID}-19},'' \emph{\BIBforeignlanguage{en}{Front. Physiol.}},
  vol.~13, p. 848172, Mar. 2022.

\bibitem{bandt_permutation_2002}
C.~Bandt and B.~Pompe, ``\BIBforeignlanguage{en}{Permutation {Entropy}: {A}
  {Natural} {Complexity} {Measure} for {Time} {Series}},''
  \emph{\BIBforeignlanguage{en}{Phys. Rev. Lett.}}, vol.~88, no.~17, p. 174102,
  Apr. 2002.

\bibitem{rostaghi_dispersion_2016}
M.~Rostaghi and H.~Azami, ``\BIBforeignlanguage{en}{Dispersion {Entropy}: {A}
  {Measure} for {Time}-{Series} {Analysis}},''
  \emph{\BIBforeignlanguage{en}{IEEE Signal Processing Letters}}, vol.~23,
  no.~5, pp. 610--614, May 2016.

\bibitem{azami_amplitude-_2018}
H.~Azami and J.~Escudero, ``\BIBforeignlanguage{en}{Amplitude- and
  {Fluctuation}-{Based} {Dispersion} {Entropy}},''
  \emph{\BIBforeignlanguage{en}{Entropy}}, vol.~20, no.~3, p. 210, Mar. 2018.

\bibitem{olofsen_permutation_2008}
E.~Olofsen, J.~Sleigh, and A.~Dahan, ``\BIBforeignlanguage{en}{Permutation
  entropy of the electroencephalogram: a measure of anaesthetic drug effect},''
  \emph{\BIBforeignlanguage{en}{British Journal of Anaesthesia}}, vol. 101,
  no.~6, pp. 810--821, Dec. 2008.

\bibitem{pincus_approximate_1991}
S.~M. Pincus, ``\BIBforeignlanguage{en}{Approximate entropy as a measure of
  system complexity.}'' \emph{\BIBforeignlanguage{en}{Proceedings of the
  National Academy of Sciences}}, vol.~88, no.~6, pp. 2297--2301, Mar. 1991.

\bibitem{caldirola_approximate_2004}
D.~Caldirola, L.~Bellodi, A.~Caumo, G.~Migliarese, and G.~Perna,
  ``\BIBforeignlanguage{en}{Approximate {Entropy} of {Respiratory} {Patterns}
  in {Panic} {Disorder}},'' \emph{\BIBforeignlanguage{en}{American Journal of
  Psychiatry}}, vol. 161, no.~1, pp. 79--87, Jan. 2004.

\bibitem{richman_physiological_2000}
J.~S. Richman and J.~R. Moorman, ``\BIBforeignlanguage{en}{Physiological
  time-series analysis using approximate entropy and sample entropy},''
  \emph{\BIBforeignlanguage{en}{American Journal of Physiology-Heart and
  Circulatory Physiology}}, vol. 278, no.~6, pp. H2039--H2049, Jun. 2000.

\bibitem{lake_sample_2002}
D.~E. Lake, J.~S. Richman, M.~P. Griffin, and J.~R. Moorman,
  ``\BIBforeignlanguage{en}{Sample entropy analysis of neonatal heart rate
  variability},'' \emph{\BIBforeignlanguage{en}{American Journal of
  Physiology-Regulatory, Integrative and Comparative Physiology}}, vol. 283,
  no.~3, pp. R789--R797, Sep. 2002.

\bibitem{weiting_chen_characterization_2007}
{Weiting Chen}, {Zhizhong Wang}, {Hongbo Xie}, and {Wangxin Yu},
  ``\BIBforeignlanguage{en}{Characterization of {Surface} {EMG} {Signal}
  {Based} on {Fuzzy} {Entropy}},'' \emph{\BIBforeignlanguage{en}{IEEE Trans.
  Neural Syst. Rehabil. Eng.}}, vol.~15, no.~2, pp. 266--272, Jun. 2007.

\bibitem{azami_multivariate_2019}
H.~Azami, A.~Fernández, and J.~Escudero,
  ``\BIBforeignlanguage{en}{Multivariate {Multiscale} {Dispersion} {Entropy} of
  {Biomedical} {Times} {Series}},'' \emph{\BIBforeignlanguage{en}{Entropy}},
  vol.~21, no.~9, p. 913, Sep. 2019.

\bibitem{morabito_multivariate_2012}
F.~C. Morabito, D.~Labate, F.~La~Foresta, A.~Bramanti, G.~Morabito, and
  I.~Palamara, ``\BIBforeignlanguage{en}{Multivariate {Multi}-{Scale}
  {Permutation} {Entropy} for {Complexity} {Analysis} of {Alzheimer}’s
  {Disease} {EEG}},'' \emph{\BIBforeignlanguage{en}{Entropy}}, vol.~14, no.~7,
  pp. 1186--1202, Jul. 2012.

\bibitem{ahmed_multivariate_2011}
M.~U. Ahmed and D.~P. Mandic, ``\BIBforeignlanguage{en}{Multivariate multiscale
  entropy: {A} tool for complexity analysis of multichannel data},''
  \emph{\BIBforeignlanguage{en}{Phys. Rev. E}}, vol.~84, no.~6, p. 061918, Dec.
  2011.

\bibitem{azami_memd-enhanced_2016}
H.~Azami, K.~Smith, and J.~Escudero, ``\BIBforeignlanguage{en}{{MEMD}-enhanced
  multivariate fuzzy entropy for the evaluation of complexity in biomedical
  signals},'' in \emph{\BIBforeignlanguage{en}{2016 38th {Annual}
  {International} {Conference} of the {IEEE} {Engineering} in {Medicine} and
  {Biology} {Society} ({EMBC})}}.\hskip 1em plus 0.5em minus 0.4em\relax
  Orlando, FL, USA: IEEE, Aug. 2016, pp. 3761--3764.

\bibitem{luo_review_2010}
S.~Luo and P.~Johnston, ``\BIBforeignlanguage{en}{A review of electrocardiogram
  filtering},'' \emph{\BIBforeignlanguage{en}{Journal of Electrocardiology}},
  p.~11, 2010.

\bibitem{dressler_awareness_2004}
O.~Dressler, G.~Schneider, G.~Stockmanns, and E.~Kochs,
  ``\BIBforeignlanguage{en}{Awareness and the {EEG} power spectrum: analysis of
  frequencies},'' \emph{\BIBforeignlanguage{en}{British Journal of
  Anaesthesia}}, vol.~93, no.~6, pp. 806--809, Dec. 2004.

\bibitem{stauss_identification_2007}
H.~M. Stauss, ``\BIBforeignlanguage{en}{Identification of blood pressure
  control mechanisms by power spectral analysis},''
  \emph{\BIBforeignlanguage{en}{Clinical and Experimental Pharmacology and
  Physiology}}, vol.~34, no.~4, pp. 362--368, Feb. 2007.

\bibitem{kaczka_assessment_1995}
D.~W. Kaczka, G.~M. Barnas, B.~Suki, and K.~R. Lutchen,
  ``\BIBforeignlanguage{en}{Assessment of time-domain analyses for estimation
  of low-frequency respiratory mechanical properties and impedance spectra},''
  \emph{\BIBforeignlanguage{en}{Ann Biomed Eng}}, vol.~23, no.~2, pp. 135--151,
  Mar. 1995.

\bibitem{diong_modeling_2007}
B.~Diong, H.~Nazeran, P.~Nava, and M.~Goldman,
  ``\BIBforeignlanguage{en}{Modeling {Human} {Respiratory} {Impedance}},''
  \emph{\BIBforeignlanguage{en}{IEEE Eng. Med. Biol. Mag.}}, vol.~26, no.~1,
  pp. 48--55, Jan. 2007.

\bibitem{gu_optimizing_2022}
H.~Gu and C.-A. Chou, ``\BIBforeignlanguage{en}{Optimizing non-uniform
  multivariate embedding for multiscale entropy analysis of complex systems},''
  \emph{\BIBforeignlanguage{en}{Biomedical Signal Processing and Control}},
  vol.~71, p. 103206, Jan. 2022.

\bibitem{xiao_variational_2021}
H.~Xiao and D.~P. Mandic, ``\BIBforeignlanguage{en}{Variational {Embedding}
  {Multiscale} {Sample} {Entropy}: {A} {Tool} for {Complexity} {Analysis} of
  {Multichannel} {Systems}},'' \emph{\BIBforeignlanguage{en}{Entropy}},
  vol.~24, no.~1, p.~26, Dec. 2021.

\bibitem{xie_cross-fuzzy_2010}
H.-B. Xie, Y.-P. Zheng, J.-Y. Guo, and X.~Chen,
  ``\BIBforeignlanguage{en}{Cross-fuzzy entropy: {A} new method to test pattern
  synchrony of bivariate time series},''
  \emph{\BIBforeignlanguage{en}{Information Sciences}}, vol. 180, no.~9, pp.
  1715--1724, May 2010.

\bibitem{shi_coupling_2015}
W.~Shi, P.~Shang, and A.~Lin, ``\BIBforeignlanguage{en}{The coupling analysis
  of stock market indices based on cross-permutation entropy},''
  \emph{\BIBforeignlanguage{en}{Nonlinear Dyn}}, vol.~79, no.~4, pp.
  2439--2447, Mar. 2015.

\bibitem{costa_multiscale_2002}
M.~Costa, A.~L. Goldberger, and C.-K. Peng,
  ``\BIBforeignlanguage{en}{Multiscale {Entropy} {Analysis} of {Complex}
  {Physiologic} {Time} {Series}},'' \emph{\BIBforeignlanguage{en}{Phys. Rev.
  Lett.}}, vol.~89, no.~6, p. 068102, Jul. 2002.

\bibitem{ahmed_multivariate_2012}
M.~U. Ahmed and D.~P. Mandic, ``\BIBforeignlanguage{en}{Multivariate
  {Multiscale} {Entropy} {Analysis}},'' \emph{\BIBforeignlanguage{en}{IEEE
  Signal Process. Lett.}}, vol.~19, no.~2, pp. 91--94, Feb. 2012.

\bibitem{valencia_refined_2009}
J.~F. Valencia, A.~Porta, M.~Vallverdu, F.~Claria, R.~Baranowski,
  E.~Orlowska-Baranowska, and P.~Caminal, ``\BIBforeignlanguage{en}{Refined
  {Multiscale} {Entropy}: {Application} to 24-h {Holter} {Recordings} of
  {Heart} {Period} {Variability} in {Healthy} and {Aortic} {Stenosis}
  {Subjects}},'' \emph{\BIBforeignlanguage{en}{IEEE Trans. Biomed. Eng.}},
  vol.~56, no.~9, pp. 2202--2213, Sep. 2009.

\bibitem{hamed_azami_coarse-graining_2018}
{Hamed Azami} and {Javier Escudero},
  ``\BIBforeignlanguage{en}{Coarse-{Graining} {Approaches} in {Univariate}
  {Multiscale} {Sample} and {Dispersion} {Entropy}},''
  \emph{\BIBforeignlanguage{en}{Entropy}}, vol.~20, no.~2, p. 138, Feb. 2018.

\bibitem{kafantaris_augmentation_2020}
E.~Kafantaris, I.~Piper, T.-Y.~M. Lo, and J.~Escudero,
  ``\BIBforeignlanguage{en}{Augmentation of {Dispersion} {Entropy} for
  {Handling} {Missing} and {Outlier} {Samples} in {Physiological} {Signal}
  {Monitoring}},'' \emph{\BIBforeignlanguage{en}{Entropy}}, vol.~22, no.~3, p.
  319, Mar. 2020.

\bibitem{costa_multiscale_2005}
M.~Costa, A.~L. Goldberger, and C.-K. Peng,
  ``\BIBforeignlanguage{en}{Multiscale entropy analysis of biological
  signals},'' \emph{\BIBforeignlanguage{en}{Physical Review E}}, vol.~71,
  no.~2, Feb. 2005.

\bibitem{fogedby_phase_1992}
H.~C. Fogedby, ``\BIBforeignlanguage{en}{On the phase space approach to
  complexity},'' \emph{\BIBforeignlanguage{en}{J Stat Phys}}, vol.~69, no. 1-2,
  pp. 411--425, Oct. 1992.

\bibitem{azami_refined_2017}
H.~Azami and J.~Escudero, ``\BIBforeignlanguage{en}{Refined composite
  multivariate generalized multiscale fuzzy entropy: {A} tool for complexity
  analysis of multichannel signals},'' \emph{\BIBforeignlanguage{en}{Physica A:
  Statistical Mechanics and its Applications}}, vol. 465, pp. 261--276, Jan.
  2017.

\bibitem{humeau-heurtier_multivariate_2016}
A.~Humeau-Heurtier, ``\BIBforeignlanguage{en}{Multivariate {Generalized}
  {Multiscale} {Entropy} {Analysis}},''
  \emph{\BIBforeignlanguage{en}{Entropy}}, vol.~18, no.~11, p. 411, Nov. 2016.

\bibitem{ichimaru_development_1999}
Y.~Ichimaru and G.~Moody, ``\BIBforeignlanguage{en}{Development of the
  polysomnographic database on {CD}‐{ROM}},''
  \emph{\BIBforeignlanguage{en}{Psychiatry and Clinical Neurosciences}},
  vol.~53, no.~2, pp. 175--177, Apr. 1999.

\bibitem{goldberger_physiobank_2000}
A.~L. Goldberger, L.~A.~N. Amaral, L.~Glass, J.~M. Hausdorff, P.~C. Ivanov,
  R.~G. Mark, J.~E. Mietus, G.~B. Moody, C.-K. Peng, and H.~E. Stanley,
  ``\BIBforeignlanguage{en}{{PhysioBank}, {PhysioToolkit}, and {PhysioNet}:
  {Components} of a {New} {Research} {Resource} for {Complex} {Physiologic}
  {Signals}},'' \emph{\BIBforeignlanguage{en}{Circulation}}, vol. 101, no.~23,
  Jun. 2000.

\bibitem{bashan_network_2012}
A.~Bashan, R.~P. Bartsch, J.~W. Kantelhardt, S.~Havlin, and P.~C. Ivanov,
  ``\BIBforeignlanguage{en}{Network physiology reveals relations between
  network topology and physiological function},''
  \emph{\BIBforeignlanguage{en}{Nature Communications}}, vol.~3, no.~1, Jan.
  2012.

\bibitem{hedges_distribution_1981}
L.~V. Hedges, ``\BIBforeignlanguage{en}{Distribution {Theory} for {Glass}'s
  {Estimator} of {Effect} size and {Related} {Estimators}},''
  \emph{\BIBforeignlanguage{en}{Journal of Educational Statistics}}, vol.~6,
  no.~2, pp. 107--128, Jun. 1981.

\bibitem{mongin_heart_2021}
D.~Mongin, C.~Chabert, D.~S. Courvoisier, J.~García-Romero, and J.~R.
  Alvero-Cruz, ``\BIBforeignlanguage{en}{Heart rate recovery to assess fitness:
  comparison of different calculation methods in a large cross-sectional
  study},'' \emph{\BIBforeignlanguage{en}{Research in Sports Medicine}}, pp.
  1--14, Jul. 2021.

\bibitem{rostaghi_fuzzy_2021}
M.~Rostaghi, M.~M. Khatibi, M.~R. Ashory, and H.~Azami,
  ``\BIBforeignlanguage{en}{Fuzzy {Dispersion} {Entropy}: {A} {Nonlinear}
  {Measure} for {Signal} {Analysis}},'' \emph{\BIBforeignlanguage{en}{IEEE
  Trans. Fuzzy Syst.}}, pp. 1--1, 2021.

\end{thebibliography}

\end{document}